\documentclass[prb,twocolumn,aps,superscriptaddress,floatfix]{revtex4}
\usepackage{amsmath}
\usepackage{amssymb}
\usepackage{graphicx}
\usepackage{color}
\usepackage[svgnames]{xcolor}
\usepackage[normalem]{ulem}
\usepackage{cancel}
\usepackage{hyperref}
\usepackage{verbatim}% Long comments

\begin{document}

\title{Ordered states in AB~bilayer graphene in SU(4)-symmetric model}

\author{A.V. Rozhkov}
\affiliation{Institute for Theoretical and Applied Electrodynamics, Russian
Academy of Sciences, 125412 Moscow, Russia}

\author{A.O. Sboychakov}
\affiliation{Institute for Theoretical and Applied Electrodynamics, Russian
Academy of Sciences, 125412 Moscow, Russia}

\author{A.L. Rakhmanov}
\affiliation{Institute for Theoretical and Applied Electrodynamics, Russian
Academy of Sciences, 125412 Moscow, Russia}

\begin{abstract}
We apply SU(4)-symmetric model to examine possible ordered states in AB
stacked bilayer graphene (AB-BLG). The Hamiltonian of the system possesses
this symmetry under certain assumptions. In such a model the multicomponent
order parameter can be presented as a
$4\times4$
matrix
$\hat{Q}$.
Using a mean field approximation, we derive a self-consistency equation for
$\hat{Q}$.
Among possible solutions of the obtained equation there are anomalous
quantum Hall states, spin, charge, spin-valley density waves, inter-layer
excitonic phases and their combinations. We argue that the proposed
approach is a useful tool for classification of the ordered states in
AB-BLG. The ordered states of the SU(4)-symmetric model demonstrate
extensive degeneracies. The degeneracies can be removed by taking into
account the neglected non-SU(4)-symmetric terms, disorder, substrate, as
well as other perturbations. The ordered states have varying sensitivity to
these factors. This suggests that, for AB-BLG, a ground state ordering type
is a non-universal property, susceptible to particulars of extrinsic
conditions.
\end{abstract}

%\pacs{Gr,HM}

\date{\today}

\maketitle

\section{Introduction}
%%%%%%%%%%%%%%%%%%%%%%%%%%%%%%%%%%%%%%%%%%%%%%%%%%%%%%%%%%
\label{intro}
%%%%%%%%%%%%%%%%%%%%%%%%%%%%%%%%%%%%%%%%%%%%%%%%%%%%%%%%%%

The electronic ordered states of graphene-based bilayer systems are studied
in a number of experimental and theoretical research papers. The reason of
such an interest is a richness of the phase diagrams
of these materials. Indeed, a number of the ordered phases were observed
and/or theoretically predicted in the bilayers. Among them are
magnetic~\cite{min_pseudo_fm2008, lang_af_hubb2012,
lemonic_rg_nemat_long2012, haritonov_afm2012, af_1princ2013,
prl2012aa_rakhmanov},
``correlated"-insulator~\cite{twist_exp_insul2018, twist_exp_sc2018,
twisted_nematicity2020sboychakov},
fractional-metal~\cite{de2022cascade, zhou2022isospin,
seiler2022quantum, sboychakov_FraM2021prb_lett, rakhmanov2023ab_FraM},
superconducting~\cite{twist_exp_sc2018, zhou2022isospin,
lemonic_rg_nemat_long2012, sboychakov2023triplet},
and
excitonic~\cite{AkzyanovAABLG2014}
states. They can be both uniform and
inhomogeneous~\cite{inhomogen2007stm_apl, Freitag20122053,
sboychakov2020phase_sep_jetp, Sboychakov_PRB2013_PS_AAgraph}.
It is also important that the free energy differences between these
numerous phases are small compared to a characteristic energy scale of the
electronic system. As a result, the question about the ``true'' ground
state hung in the air since the answer strongly depends on the details of a
particular problem, e.g., substrate, sample quality, doping, etc. Moreover,
externally-driven cascades of the phase transitions were observed
experimentally~\cite{de2022cascade, seiler2022quantum},
predicted, and investigated
theoretically~\cite{vafek2010interacting,cvetkovich2012bilayer_orders_theor,
szabo2022competing, chichinadze2022cascade,dong2023isospin,koh2024symmetry,
raines2024Letters,raines2024unconventional}
in Bernal or AB-stacked bilayer graphene (AB-BLG).

In this situation, it is useful to develop a reasonably general
classification scheme that allows one to itemize possible ordered phases in
a specific material. In Ref.~\onlinecite{roy2013classification} such a
classification scheme was formulated being based on the symmetries of the
non-interacting AB-BLG model. In
Ref.~\onlinecite{szabo2021extended}
this classification was extended using a Hubbard-like model. An alternative
approach to the problem was developed in
Ref.~\onlinecite{Nandkishore2010b},
where a symmetry-based analysis of competition between different
non-superconducting gapped states in AB-BLG was proposed. Such an approach
looks more compact and convenient. The mean field ordered states were
classified with the help of a ``hidden'' SU(4) symmetry. [The emergence of
such an extensive symmetry group in the model is a consequence of the SU(2)
spin-rotation invariance enhanced by the well-known valley degeneracy, a
common
feature~\cite{fischer2011su4_graphene_magnetic, chichinadze2022tblg_su4,
aa2023su4_rozhkovetal}
for various graphene-based systems.] The introduced multicomponent order
parameter can be presented as a
$4\times4$
matrix
$\hat{Q}$,
which satisfies a self-consistency equation. The solutions to the
self-consistency equation can be grouped according the value of a
topological invariant defined in Ref.~\onlinecite{Nandkishore2010b}
under assumption of the matrix $\hat{Q}$ being Hermitian.

Similar SU(4)-symmetric model for AA-stacked bilayer graphene (AA-BLG) has
been proposed and studied in our
paper Ref.~\onlinecite{aa2023su4_rozhkovetal}
using a mean field approach. We argued that the matrix order parameter
$\hat{Q}$
does not have to be Hermitian, and the self-consistency equation allows for
non-Hermitian solutions as well.

Given the success of SU(4)-symmetric model for analysis of the AA-BLG
ordered phases and certain inconsistencies between the conclusions of
Refs.~\onlinecite{Nandkishore2010b}
and~\onlinecite{aa2023su4_rozhkovetal},
not to mention an obvious fact that AB-BLG is much better explored
experimentally, we believe it is warranted to re-examine the SU(4)-symmetric
model for AB-BLG. This is the main goal of this paper.

In our investigation, we identify two types of interaction contributions,
direct and umklapp, that are consistent with the SU(4) group. This is to be
contrasted with
Ref.~\onlinecite{Nandkishore2010b},
where a simpler Hamiltonian, with a single coupling constant, was employed.
The SU(4)-invariant Hamiltonian is then analyzed within the mean field
approach, and a self-consistency equation for the matrix order parameter
$\hat{Q}$
is derived. Curiously, the self-consistency equation for the AB-BLG is
mathematically identical to that for the AA-BLG obtained in
Ref.~\onlinecite{aa2023su4_rozhkovetal}.
However, because of dissimilar crystal structures, the AB-BLG effective
coupling constants differ from the coupling constants for the AA-BLG.

Here, not only the matrix self-consistency equation generalizing the
equation of
Ref.~\onlinecite{Nandkishore2010b}
is derived but a broader set of solutions is also found. Specifically, in
addition to purely Hermitian
$\hat{Q}$,
non-Hermitian solutions are explicitly constructed and discussed. Beside
this, the topological classification of the order
parameters~\cite{Nandkishore2010b, aa2023su4_rozhkovetal}
is extended to some non-Hermitian solutions. A number of phases, such as
spin and charge-density wave orders, inter-layer excitonic orders, as well
as others, can be described by the proposed mean field approach.

The paper is organized as follows. In
Sec.~\ref{sect::TB},
we derive an approximate SU(4)-symmetric Hamiltonian for the AB-BLG.
In Sec.~\ref{sect::MF},
we apply mean-field approximation and obtain a self-consistency equation
for the multicomponent order parameter
$\hat{Q}$.
Possible types of the ordered states in the AB-BLG are analyzed
and the obtained results are discussed in
Sec.~\ref{sect::MF_states}.
In
Sec.~\ref{sect::discuss}
we summarize the obtained results.

%%%%%%%%%%%%%%%%%%%%%%%%%%%%%%%%%%%%%%%%%%%%%%%%%%
\section{Tight-binding SU(4) symmetric Hamiltonian for AB-BLG}
%%%%%%%%%%%%%%%%%%%%%%%%%%%%%%%%%%%%%%%%%%%%%%%%%%%%%%%%%%
\label{sect::TB}
%%%%%%%%%%%%%%%%%%%%%%%%%%%%%%%%%%%%%%%%%%%%%%%%%%%%%%%%%%

\subsection{Single-electron Hamiltonian}
%%%%%%%%%%%%%%%%%%%%%%%%%%%%%%%%%%%%%%%%%%%%%%%%%%%%%%%%%%
\label{subsect::4Bands}
%%%%%%%%%%%%%%%%%%%%%%%%%%%%%%%%%%%%%%%%%%%%%%%%%%%%%%%%%%

The unit cell of the AB-BLG contains four carbon atoms located in two
different layers and two different sublattices $A$ and $B$. The commonly
used tight-binding
Hamiltonian~\cite{ourBLGreview2016}
for AB-BLG reads as
\begin{equation}
%%%%%%%%%%%%%%%%%%%%%%%%%%%%%%
\label{H_0_tightBinding}
%%%%%%%%%%%%%%%%%%%%%%%%%%%%%%
\hat{H}_0=\left(
            \begin{array}{cccc}
              0 & 0 & -tf(\mathbf{k}) & 0 \\
              0 & 0 & t_0 & -tf(\mathbf{k}) \\
              -tf^*(\mathbf{k}) & t_0 & 0 & 0 \\
              0 & -tf^*(\mathbf{k}) & 0 & 0 \\
            \end{array}
          \right),
\end{equation}
where
$t \approx 2.7$\,eV
is the intralayer nearest-neighbor hoping amplitude,
$t_0 \approx 0.35$\,eV
is the hoping amplitude between nearest (`dimer') atoms in different layers,
\begin{equation}
%%%%%%%%%%%%%%%%%%%%%%%%%%%%%%
\label{f(k)}
%%%%%%%%%%%%%%%%%%%%%%%%%%%%%%
f(\mathbf{k})
=
1\!+\!2\exp{\!\left(\frac{3ia_0k_x}{2}\right)}
\cos \left( \frac{\sqrt{3}a_0k_y}{2} \right),
\end{equation}
and
$a_0 = 1.42$\,\AA\,
is the in-plane carbon-carbon distance. The wave functions of this
Hamiltonian are bi-spinors
\begin{equation}
%%%%%%%%%%%%%%%%%%%%%%%%%%%%%%
\label{Psi_start}
%%%%%%%%%%%%%%%%%%%%%%%%%%%%%%
\hat{\Psi}_{\mathbf{k}\sigma}
=
(d_{\mathbf{k}1A\sigma},d_{\mathbf{k}2A\sigma},d_{\mathbf{k}1B\sigma},d_{\mathbf{k}2B\sigma})^T,
\end{equation}
whose components
$d_{\mathbf{k}la\sigma}$
are the second-quantization operators for electrons with momentum
$\mathbf{k}$,
in the layer
$l=1,2$,
on the sublattice
$a=A,B$,
and with spin projection
$\sigma=\uparrow,\downarrow$.
We neglect here the effect of the trigonal warping arising if we take into
account a non-dimer interlayer hoping.

The spectrum of the
Hamiltonian~\eqref{H_0_tightBinding}
consists of four bands
$\varepsilon^{(s)}(\mathbf{k})$,
where the band index is
$s=1,2,3,4$.
Using
$\varepsilon^{(s)}$,
one can formally diagonalize the
Hamiltonian~\eqref{H_0_tightBinding}
\begin{equation}
%%%%%%%%%%%%%%%%%%%%%%%%%%%%%%
\label{H_0_diag}
%%%%%%%%%%%%%%%%%%%%%%%%%%%%%%
\hat{H}_0
=
\sum_{\mathbf{k},s,\sigma}
	\varepsilon^{(s)}(\mathbf{k})
	\gamma^\dag_{\mathbf{k}s\sigma}
	\gamma^{\vphantom{\dag}}_{\mathbf{k}s\sigma}.
\end{equation}
In this formula
$\gamma_{\mathbf{k}s\sigma}$
are band operators. Although, explicit expressions for
$\varepsilon^{(s)}(\mathbf{k})$
are
known~\cite{ourBLGreview2016},
they are too complicated for our purposes. Instead, we will use the
following approximate formulas
\begin{eqnarray}
%%%%%%%%%%%%%%%%%%%%%%%%%%%%%%
\label{elec_bands}
%%%%%%%%%%%%%%%%%%%%%%%%%%%%%%
\varepsilon^{(1,4)}(\mathbf{q})
=
\pm \left[t_0+\varepsilon_0(\mathbf{q})\right],
\quad
\varepsilon^{(2,3)}(\mathbf{q})
=
\pm \varepsilon_0(\mathbf{q}),
\end{eqnarray}
which are valid when
$\varepsilon_0< t_0$
or
$|\mathbf{q}|<q_0$,
where
$q_0=2t_0/3ta_0 \approx 0.08 a_0^{-1}$,
and
$\mathbf{q}=\mathbf{k}-\mathbf{K}_\xi$
is the momentum measured relative Dirac point
$\mathbf{K_{\xi}}$
located at
\begin{eqnarray}
\mathbf{K_{\xi}}=\frac{2\pi}{3\sqrt{3} a_0} (\sqrt{3}, \xi ),
\quad
\text{where}
\quad
\xi=\pm 1
\end{eqnarray}
is the Dirac cone number or valley index, and
\begin{eqnarray}
\varepsilon_0(\mathbf{q})
\approx
\frac{9t^2}{4t_0}a_0^2|\mathbf{q}|^2=t_0\frac{|\mathbf{q}|^2}{q_0^2}.
\end{eqnarray}
Equations~(\ref{elec_bands})
demonstrate that bands
$s=2$
and
$s=3$
touch each other at two Dirac points, forming (at zero doping) two Fermi
points. As a result the ground state of AB-BLG spontaneously breaks
inversion symmetry for arbitrarily weak electron-electron interactions.
Note that for the realistic values of the electron-electron interaction in
AB-BLG, the effect of the trigonal warping is weak, and it cannot eliminate
the instability of the Fermi liquid toward the spontaneous symmetry
breaking~\cite{PhysRevB.81.041402}.
Note also that the touching of the bands in the AB-BLG is quadratic in
contrast to the linear crossing of the bands in the AA-BLG and the linear
Dirac spectrum of monolayer graphene.

As usual for graphene-based systems, it is convenient to define the Dirac
cone index $\xi$ as a valley quantum number characterizing single-electron
states. We assign an electron with momentum
$\mathbf{k}$
to the valley
$\mathbf{K}_\xi$
if
$|\mathbf{q}|<q_0$.
It is easy to check that
$q_0<|\mathbf{K}_{+1}-\mathbf{K}_{-1}|/2 \approx 1.21 a_0^{-1}$
since
$t_0/t<1$.
This guarantees that the valleys do not overlap (a state with a given
momentum never belongs to both valleys at once). As for the electron states
that do not reside in any of the valleys, they are discarded since their
contribution to the low-energy physics is insignificant.

It follows from
Eqs.~\eqref{elec_bands}
that
$|\varepsilon^{(2,3)}(q_0)|\leq |\varepsilon^{(1,4)}(\mathbf{q})|$
and we can exclude bands
$s=1,4$
from our low-energy treatment. This reduces the formulated above
tight-binding approach to a generally accepted two-bands effective
model~\cite{ourBLGreview2016}
for AB-BLG. In this case the
Hamiltonian~\eqref{H_0_diag}
reads
\begin{eqnarray}
%%%%%%%%%%%%%%%%%%%%%%%%%%%%%%
\label{H_0_eff_2B}
%%%%%%%%%%%%%%%%%%%%%%%%%%%%%%
\hat{H}_0^{\textrm{eff}}
=
\!\sum_{{\bf q} ,\xi,\sigma}\!
	\left(
		\gamma_{\mathbf{q}2\xi\sigma}^\dag,
		\gamma_{\mathbf{q}3\xi\sigma}^\dag
	\right)\!
	\left(
		\begin{matrix}
			\varepsilon_0 ({\bf q}) & 0 \\
			0 & -\varepsilon_0 ({\bf q}) \\
		\end{matrix}
	\right)\!
	\left(
		\begin{matrix}
			\gamma_{\mathbf{q}2\xi\sigma}^{\vphantom{\dag}} \\
			\gamma_{\mathbf{q}3\xi\sigma}^{\vphantom{\dag}}  \\
		\end{matrix}
	\right)\!.
%	\varepsilon_0(\mathbf{q})
%	\left(
%		\gamma^\dag_{\mathbf{q}2\xi\sigma}
%		\gamma_{\mathbf{q}2\xi\sigma}
%		-
%		\gamma^\dag_{\mathbf{q}3\xi\sigma}
%		\gamma_{\mathbf{q}3\xi\sigma}
%	\right).
\end{eqnarray}
In the lowest-order approximation in powers of
$q/q_0 < 1$,
the relations between the valley-specific band operators
$\gamma_{{\bf q}s\xi \sigma}$
and valley-specific lattice operators
$d_{{\bf q}l a \xi \sigma}$
can be presented in the form
\begin{eqnarray}
%%%%%%%%%%%%%%%%%%%%%%%%%%%%%%
\label{d_q0LoEn_g}
%%%%%%%%%%%%%%%%%%%%%%%%%%%%%%
\nonumber
d_{\mathbf{q}1A\xi\sigma}
&=&
\frac{e^{ i\xi\phi_\mathbf{q}}}{\sqrt{2}}
\left(
	- \gamma_{\mathbf{q}2\xi\sigma} + \xi \gamma_{\mathbf{q}3\xi\sigma}
\right),
\\
d_{\mathbf{q}2B\xi\sigma}
&=&
\frac{e^{ - i\xi\phi_\mathbf{q}}}{\sqrt{2}}
\left(
	\gamma_{\mathbf{q}2\xi\sigma} + \xi \gamma_{\mathbf{q}3\xi\sigma}
\right),
\end{eqnarray}
where
$\exp(i \phi_\mathbf{q})=-(iq_x + q_y)/|{\bf q}|$,
and
$\xi = -1$
($\xi = 1$)
for
${\bf K}_1$
(for
${\bf K}_2$).
Deriving these compact expressions we employed the fact that any band
operator is defined up to a (undetermined) phase factor, which enables us
to fix this phase based on convenience. Note also that in our approximation
the dimer sites do not participate in the low-energy electron physics.

In
expression~(\ref{H_0_eff_2B}),
the band energies are independent of the spin and valley indices, while
different spin projections and different valley indices correspond to
different electronic states. We see that the single-electron Hamiltonian
$\hat{H}^{\rm eff}_0$
is SU(4) symmetric in the multi-index space
$m=(\xi,\sigma)$.
That is,
$\hat{H}_0^{\rm eff}$
is invariant under Bogolyubov transformation
\begin{eqnarray}
%%%%%%%%%%%%%%%%%%%%%%%%%%%%%%%%%%%%%%%%%%%%%%%%%%
\label{invariance}
%%%%%%%%%%%%%%%%%%%%%%%%%%%%%%%%%%%%%%%%%%%%%%%%%%
\gamma_{\mathbf{q}sm}'=\sum_{m'}{u_{mm'}\gamma_{\mathbf{q}sm'}},
\end{eqnarray}
where
$u_{mm'}$
are matrix elements of a
$4\times 4$
unitary matrix.

\subsection{Effective interaction Hamiltonian}
%%%%%%%%%%%%%%%%%%%%%%%%%%%%%%%%%%%%%%%%%%%%%%%%%%%%%%%%%%
\label{subsect::Interaction}
%%%%%%%%%%%%%%%%%%%%%%%%%%%%%%
%%%%%%%%%%%%%%%%%%%%%%%%%%%%%%%%%%%%%%%%%%%%%%%%%%%%%%%%%%

The interaction part of the Hamiltonian
$\hat{H}_{\textrm{int}}$
is assumed to be spin-independent density-density
type~\cite{aa2023su4_rozhkovetal}
\begin{equation}
%%%%%%%%%%%%%%%%%%%%%%%%%%%%%%
\label{Int_General}
%%%%%%%%%%%%%%%%%%%%%%%%%%%%%%
\hat{H}_{\textrm{int}}
=
\frac{1}{2N_c}\sum_{\mathbf{k}ll'}
	V^{ll'}_{\mathbf{k}}
	\hat{\rho}_{\mathbf{k}l}
	\hat{\rho}_{-\mathbf{k}l'}.
\end{equation}
In this formula
$N_c$
is the number of unit cells in the system,
$\mathbf{k}$
is a transferred momentum,
$V^{ll'}_{\mathbf{k}}$
is the Fourier transform of the potential energy
$V^{ll'}(\mathbf{r})$
of the interaction between an electron in the layer $l$ and another
electron in the layer $l'$, and
$\mathbf{r}$
is a distance between these electrons. The symbol
$\hat{\rho}_{\mathbf{k}l}$
denotes the Fourier component of a single-site density operator defined for
elementary cell
$\mathbf{n}$
as
$\hat{\rho}^{\vphantom{\dag}}_{\mathbf{n}1}
=
\sum_{\sigma}d^\dag_{\mathbf{n}1A\sigma}
d^{\vphantom{\dag}}_{\mathbf{n}1A\sigma}$
and
$\hat{\rho}^{\vphantom{\dag}}_{\mathbf{n}2}
=
\sum_{\sigma}d^\dag_{\mathbf{n}2B\sigma}
d^{\vphantom{\dag}}_{\mathbf{n}2B\sigma}$.
\begin{figure}
\includegraphics[width=0.99\columnwidth]{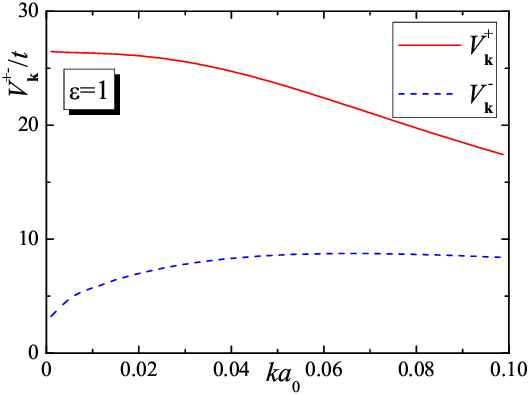}
\\
\includegraphics[width=0.99\columnwidth]{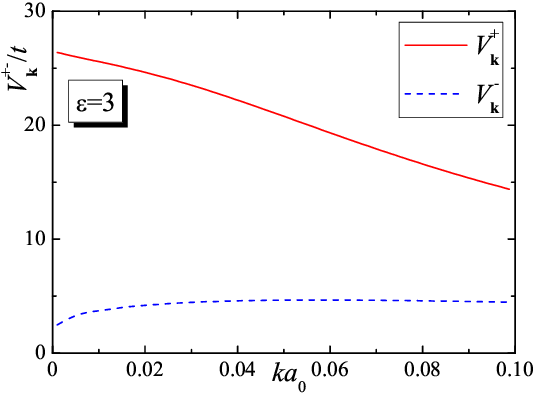}
\caption{Effective interactions
$V^\pm_{\bf k}$
calculated using the RPA approach for AB-BLG, as in
Ref.~\onlinecite{sboychakov2023triplet}.
The panels show the data for two different values of the dielectric
constants of the media surrounding the AB-BLG sample:
$\varepsilon = 1$
for the top panel and
$\varepsilon = 3$
for the bottom panel. Solid (red) curves on both panels represent
$V^+_{\bf k}$
as function of transferred momentum
${\bf k}$
for
$k = |{\bf k}| < q_0$.
Dashed (blue) curves represent
$V^-_{\bf k}$.
Dependence on the direction of
${\bf k}$
is weak and may be neglected. For both values of the dielectric constant we
can clearly observe that
$V^-_{\bf k} < V^+_{\bf k}$.
%%%%%%%%%%%%%%%%%%%%%%%%%%%%%%%%%%%%%%%%%%%%%%%%%%
\label{interaction}
%%%%%%%%%%%%%%%%%%%%%%%%%%%%%%%%%%%%%%%%%%%%%%%%%%
}
\end{figure}

We assume that the interaction is weaker at large transferred momenta and
stronger at smaller ones. In particular,
$|V^{ll'}_{{\bf K}_\xi}| < V^{ll'}_{\bf k}$
for
$|{\bf k}| < q_0$.
In such a situation, one can neglect the so-called backscattering and
consider only the case of small
$|{\bf k}|<q_0$,
which guarantees that the scattered electrons remain in their valleys.
Thus, we can define `chiral' components of the density operators
\begin{eqnarray}
%%%%%%%%%%%%%%%%%%%%%%%%%%%%%%
\label{ro_xi1}
%%%%%%%%%%%%%%%%%%%%%%%%%%%%%%
\hat{\rho}\,^\xi_{\mathbf{k}1}\
=
\sum_{\mathbf{q}\sigma}
	d^\dag_{\mathbf{q}1A\xi\sigma}
	d^{\vphantom{\dag}}_{\mathbf{q}+\mathbf{k}1A\xi\sigma},
\\
%%%%%%%%%%%%%%%%%%%%%%%%%%%%%%
\label{ro_xi2}
%%%%%%%%%%%%%%%%%%%%%%%%%%%%%%
\hat{\rho}\,^\xi_{\mathbf{k}2}
=
\sum_{\mathbf{q}\sigma}
	d^\dag_{\mathbf{q}2B\xi\sigma}
	d^{\vphantom{\dag}}_{\mathbf{q}+\mathbf{k}2B\xi\sigma},
\end{eqnarray}
and approximate the interaction as
\begin{equation}
%%%%%%%%%%%%%%%%%%%%%%%%%%%%%%
\label{Int_low_E}
%%%%%%%%%%%%%%%%%%%%%%%%%%%%%%
\hat{H}_{\textrm{int}}
\approx
\frac{1}{2N_c}\!\sum_{{\bf k}\xi\xi'll'}\!\!
	V^{ll'}_{\mathbf{k}}
	\hat{\rho}\,^\xi_{\mathbf{k}l}
	\hat{\rho}\,^{\xi'}_{-\mathbf{k}l'}.
\end{equation}
Finally, disregarding any layer asymmetry, we assume that
$V^{11}_{\mathbf{k}}=V^{22}_{\mathbf{k}}$,
and
$V^{12}_{\mathbf{k}}=V^{21}_{\mathbf{k}}<V^{11}_{\mathbf{k}}$.

Introducing linear combinations of the interaction potentials
$V^{\pm}_{\mathbf{k}}=V^{11}_{\mathbf{k}}\pm V^{12}_{\mathbf{k}}$,
we express
Hamiltonian~\eqref{Int_low_E}
in the following form
\begin{equation}
%%%%%%%%%%%%%%%%%%%%%%%%%%%%%%
\label{Int_low_E2}
%%%%%%%%%%%%%%%%%%%%%%%%%%%%%%
\hat{H}_{\textrm{int}}
\approx
\frac{1}{4N_c}\sum_{\bf k}
	\left(
		V^{+}_{\mathbf{k}}
		\hat{\rho}^{\vphantom{+}}_{\mathbf{k}}
		\hat{\rho}^{\vphantom{+}}_{-\mathbf{k}}
		+		
		V^{-}_{\mathbf{k}}
		\hat{\chi}^{\vphantom{+}}_{\mathbf{k}}
		\hat{\chi}^{\vphantom{+}}_{-\mathbf{k}}
	\right)\!,
\end{equation}
where
\begin{eqnarray}
%%%%%%%%%%%%%%%%%%%%%%%%%%%%%%
\label{ro_chi}
%%%%%%%%%%%%%%%%%%%%%%%%%%%%%%
\hat{\rho}_{\mathbf{k}}
=
\sum_\xi\left(
	\hat{\rho}\,^\xi_{\mathbf{k}1}+\hat{\rho}\,^\xi_{\mathbf{k}2}
\right),
\quad
\hat{\chi}_{\mathbf{k}}
=
\sum_\xi\left(
	\hat{\rho}\,^\xi_{\mathbf{k}1}-\hat{\rho}\,^\xi_{\mathbf{k}2}
\right).
\end{eqnarray}
We can identify
$\hat{\rho}_{\mathbf{k}}$
as electron density operator for the bilayer, while
$\hat{\chi}_{\mathbf{k}}$
can be viewed as a dipole-moment density operator. Consequently,
$V^+_{\bf k}$
represents the interaction between electric charges, while
$V^-_{\bf k}$
describes dipole-dipole interaction.

Using
Eqs.~\eqref{d_q0LoEn_g},
\eqref{ro_xi1}
and~\eqref{ro_xi2}
we can express
$\hat{\rho}_{\bf k}$
in terms of the band operators
%\begin{widetext}
\begin{eqnarray}
%%%%%%%%%%%%%%%%%%%%%%%%%%%%%%
\label{rho+final}
%%%%%%%%%%%%%%%%%%%%%%%%%%%%%%
\hat{\rho}_{\mathbf{k}}
=
\sum_{\mathbf{q}m}
	\left[
		\cos{\phi_{\mathbf{qk}}}
		\left(
			\gamma^\dag_{\mathbf{q}2m}
			\gamma^{\vphantom{\dag}}_{\mathbf{q+k}2m}
			+
			\gamma^\dag_{\mathbf{q}3 m}
			\gamma^{\vphantom{\dag}}_{\mathbf{q+k}3m}
		\right)
	\right.
\\
\nonumber
	\left.
		+i\sin{\phi_{\mathbf{qk}}}
		\left(
			\gamma^\dag_{\mathbf{q}2m}
			\gamma^{\vphantom{\dag}}_{\mathbf{q+k}3m}
			+
			\gamma^\dag_{\mathbf{q}3m}
			\gamma^{\vphantom{\dag}}_{\mathbf{q+k}2m}
		\right)
	\right],
\end{eqnarray}
where
$\phi_{\mathbf{q}\mathbf{k}}
=
\phi_{\mathbf{q}} - \phi_{\mathbf{q}+\mathbf{k}}$.
This form explicitly demonstrates that
$\hat{\rho}_{\bf k}$
is SU(4)-invariant operator.

At the same time, the dipole-momentum density operator does not have the
same invariance, as one can see from the formula
\begin{eqnarray}
%%%%%%%%%%%%%%%%%%%%%%%%%%%%%%
\label{chi+final}
%%%%%%%%%%%%%%%%%%%%%%%%%%%%%%
\hat{\chi}_{\mathbf{k}}
\!=\!
-\!\!\sum_{\mathbf{q}\xi\sigma}
	\xi\!
	\left[
		i  \sin{\phi_{\mathbf{qk}}} \!
		\left(
			\gamma^\dag_{\mathbf{q}2\xi \sigma}
			\gamma^{\vphantom{\dag}}_{\mathbf{q+k}2\xi \sigma}
			\!+
			\gamma^\dag_{\mathbf{q}3 \xi \sigma}
			\gamma^{\vphantom{\dag}}_{\mathbf{q+k}3 \xi \sigma}
		\right)
	\right.
\\
\nonumber
	\left.
		+\cos {\phi_{\mathbf{qk}}}
		\left(
			\gamma^\dag_{\mathbf{q}2 \xi \sigma }
			\gamma^{\vphantom{\dag}}_{\mathbf{q+k}3 \xi \sigma}
			+
			\gamma^\dag_{\mathbf{q}3 \xi \sigma }
			\gamma^{\vphantom{\dag}}_{\mathbf{q+k}2 \xi \sigma}
		\right)
	\right].
\end{eqnarray}
The sign-oscillating factor
$\xi=\pm1$
in this relation explicitly spoils the invariance.

Following
Ref.~\onlinecite{Nandkishore2010b},
we disregard dipole-dipole interaction
$V^-_\mathbf{q}$
in the
Hamiltonian~\eqref{Int_low_E2}
since
$V^+_\mathbf{q}>V^-_\mathbf{q}$.
To explain this simplification, let us examine
Fig.~\ref{interaction}.
Its two panels present the AB-BLG effective interaction calculated within
the random-phase approximation (RPA)
framework~\cite{sboychakov2023triplet}.
We see that for both values of the dielectric constant, the latter
inequality remains valid in the relevant transferred momenta window
$|{\bf k}| < q_0$.

Thus, the effective Hamiltonian can be simplified as
\begin{equation}
%%%%%%%%%%%%%%%%%%%%%%%%%%%%%%
\label{H_eff_fin}
%%%%%%%%%%%%%%%%%%%%%%%%%%%%%%
\hat{H}_{\textrm{int}}
\approx
\frac{1}{4N_c}\sum_{{\bf k} }
	V^{+}_{\mathbf{k}}
	\hat{\rho}^{\vphantom{+}}_{\mathbf{k}}
	\hat{\rho}^{\vphantom{+}}_{-\mathbf{k}}.
\end{equation}
The obtained interaction Hamiltonian is SU(4) invariant. Its SU(4)
symmetry, as well as the symmetry of the single-electron
Hamiltonian~(\ref{H_0_eff_2B}),
follows from the fact that the band energies
$\varepsilon_0 (\mathbf{q})$
and the matrix elements
$V_{\mathbf{k}}^{ll'}$
are independent of the multi-index $m$, while bilinears
\begin{equation*}
\sum_m \gamma^\dag_{\mathbf{q}sm}\gamma^{\vphantom{\dag}}_{\mathbf{q'}s'm}
\end{equation*}
in
Eq.~(\ref{H_0_eff_2B})
and~(\ref{rho+final})
remain unchanged under the action of the SU(4) Bogolyubov
transformation~(\ref{invariance}).

Note that according to Fig.~\ref{interaction}
at small transfer momentum interaction
$V^{+}_{\mathbf{k}}$
exceeds the value of
25$t \approx 75$\,eV.
In particular, for graphene and graphene-based materials, this is much
greater than estimates for the Hubbard-like on-site repulsion $U$, a known
non-SU(4)-symmetric contribution. At larger
$k = |{\bf k}|$,
the coupling
$V^{ll’}_{\mathbf{k}}$
decreases approximately as $1/k$ (see Fig.~2 of
Ref.~\onlinecite{sboychakov2023triplet},
where the dependence of
$V^{ll’}_{\mathbf{k}}$
on $k$ is shown in a wider $k$ range), which makes the backscattering
smaller than the forward scattering. These two arguments justifies the use
of the SU(4)-symmetric
interaction~\eqref{H_eff_fin},
at least, as a crude approximation. This issue is additionally discussed in
Sec.~\ref{sect::discuss}.

\section{Mean field approach}
%%%%%%%%%%%%%%%%%%%%%%%%%%%%%%%%%%%%%%%%%%%%%%%%%%
\label{sect::MF}
%%%%%%%%%%%%%%%%%%%%%%%%%%%%%%%%%%%%%%%%%%%%%%%%%%

Now we apply the mean field approach to the obtained effective
Hamiltonian~\eqref{H_eff_fin}.
Below, only fully gapped ordered states will be investigated (this
restriction excludes nematic
phases~\cite{vafek_nemat_rg2010}).
To open a gap in the single-particle spectrum, the order parameter must include
off-diagonal elements in the matrix for the two-band
Hamiltonian~(\ref{H_0_eff_2B}).
Thus, in ordered phases inter-band average of the form
$\langle
\gamma^\dag_{\mathbf{q}3m} \gamma^{\vphantom{\dag}}_{\mathbf{q}2m'}
\rangle$
becomes finite. We will use this ``anomalous'' average to perform the mean
field decoupling. (While many
workers~\cite{Nandkishore2010b, vafek_nemat_rg2010, gorbar2012order_compet}
prefer to introduce order parameters in terms of symmetry-breaking averages of
lattice operators
$\langle d^\dag d \rangle$,
our definition is perfectly equivalent to that.)

Keeping in mind that
$V^{+}_{\mathbf{p}-\mathbf{q}}=V^{+}_{\mathbf{q}-\mathbf{p}}$
we derive
\begin{widetext}
\begin{eqnarray}
%%%%%%%%%%%%%%%%%%%%%%%%%%%%%%
\label{Dir_MF}
%%%%%%%%%%%%%%%%%%%%%%%%%%%%%%
\hat{H}_{\textrm{int}}^{\textrm{MF}}
=
-\frac{1}{2N_c}\sum_{\mathbf{p}\mathbf{q}mm'}
	V^+_{\mathbf{p}-\mathbf{q}}
	\langle
		%\Theta_{\mathbf{q}mm'}
		\gamma^\dag_{\mathbf{p}3m}
		\gamma^{\vphantom{\dag}}_{\mathbf{p}2m'}
	\rangle
	\left[
%		\cos^2{(\alpha_{\mathbf{qp}})}
		\cos^2{( \phi_{\bf q} - \phi_{\bf p} )}
		\gamma^\dag_{\mathbf{q}2m'}
		\gamma^{\vphantom{\dag}}_{\mathbf{q}3m}
%\Theta^\dag_{\mathbf{p}m'm}
%\right.
%\\
%\nonumber
%\left.
+
		\sin^2{( \phi_{\bf q} - \phi_{\bf p} )}
		\gamma^\dag_{\mathbf{q}3m'}
		\gamma^{\vphantom{\dag}}_{\mathbf{q}2m}
%		\Theta_{\mathbf{p}m'm}
%\sin^2{(\alpha_{\mathbf{qp}})}
\right] + {\rm H.c.}
\end{eqnarray}
\end{widetext}
To make the mean-field formulas more transparent, we introduce the
$4\times 4$
operator-valued matrix
$\hat{\Theta}_\mathbf{q}$
whose matrix elements are
\begin{equation}
%%%%%%%%%%%%%%%%%%%%%%%%%%%%%%
\label{Theta}
%%%%%%%%%%%%%%%%%%%%%%%%%%%%%%
\Theta_{\mathbf{q}mm'}^{\vphantom{\dag}}
=
\gamma^\dag_{\mathbf{q}3m} \gamma^{\vphantom{\dag}}_{\mathbf{q}2m'}.
\end{equation}
As a result, we obtain a compact expression for the interaction Hamiltonian
in the mean-field approach
\begin{eqnarray}
%%%%%%%%%%%%%%%%%%%%%%%%%%%%%%
\label{MeanF_interact}
%%%%%%%%%%%%%%%%%%%%%%%%%%%%%%
\hat{H}_{\textrm{int}}^{\textrm{MF}}
&=&
\hat{H}_{\textrm{dir}}^{\textrm{MF}}+\hat{H}_{\textrm{um}}^{\textrm{MF}},
\\
\nonumber
\hat{H}_{\textrm{dir}}^{\textrm{MF}}
&=&
-\frac{1}{2N_c}
\sum_{\mathbf{q}\mathbf{p}}
	V^{\rm dir}_{\mathbf{q}\mathbf{p}}
	\textrm{Tr}\!\left(
		\langle\hat{\Theta}_\mathbf{p}^\dag\rangle
		\hat{\Theta}_\mathbf{q}
		\!+\!
		\langle\hat{\Theta}_\mathbf{p}\rangle
		\hat{\Theta}^\dag_\mathbf{q}
	\right),
\\
\nonumber
\hat{H}_{\textrm{um}}^{\textrm{MF}}
&=&
-\frac{1}{2N_c}
\sum_{\mathbf{q}\mathbf{p}}
	V^{\rm um}_{\mathbf{q}\mathbf{p}}
	\textrm{Tr}\!\left(
		\langle\hat{\Theta}_\mathbf{p}\rangle
		\hat{\Theta}_\mathbf{q}
		\!+\!
		\langle\hat{\Theta}^\dag_\mathbf{p}\rangle
		\hat{\Theta}^\dag_\mathbf{q}
	\right).
\end{eqnarray}
Here
\begin{eqnarray}
%%%%%%%%%%%%%%%%%%%%%%%%%%%%%%
\label{V_sin_cos}
%%%%%%%%%%%%%%%%%%%%%%%%%%%%%%
\nonumber
V^{\rm dir}_{\mathbf{q}\mathbf{p}}
&=&
V^+_{\mathbf{q}-\mathbf{p}}\cos^2{(\phi_{\bf q} - \phi_{\bf p})},
\\
V^{\rm um}_{\mathbf{q}\mathbf{p}}
&=&
V^+_{\mathbf{q}-\mathbf{p}}\sin^2{(\phi_{\bf q} - \phi_{\bf p})},
\end{eqnarray}
are direct (`dir') and umklapp (`um') interactions. For RPA interaction,
the function
$V^+_{\mathbf{q}}$
is finite for any $\mathbf{q}$ due to the finite density of states at the
Fermi level. This allows us to approximate
Eqs.~\eqref{MeanF_interact}
replacing the functions $V^{\rm dir}_{\mathbf{q}\mathbf{p}}$ and $V^{\rm um}_{\mathbf{q}\mathbf{p}}$ by their mean values in the range $|\mathbf{q}|<q_0$,
\begin{eqnarray}
%%%%%%%%%%%%%%%%%%%%%%%%%%%%%%%%%%%%%%%%%%%%%%%%%%
\label{average_interaction}
%%%%%%%%%%%%%%%%%%%%%%%%%%%%%%%%%%%%%%%%%%%%%%%%%%
\bar{V}_{\rm dir, um}
=
\frac{1}{\pi^2 q^4_0}
\int_{|{\bf q}|<q_0\atop |{\bf p}| < q_0}
	d^2 {\bf q} d^2 {\bf p} \, V^{\rm dir, um}_{{\bf q}{\bf p}}.
\end{eqnarray}
In essence, this formula is a familiar theoretical device common to
numerous BCS-like mean field schemes: detailed interaction function, which
is often unknown, is replaced by an effective coupling constant. For RPA
interaction at
$\varepsilon = 1$,
see
Fig.~\ref{interaction}(top),
we estimate
\begin{eqnarray}
%%%%%%%%%%%%%%%%%%%%%%%%%%%%%%%%%%%%%%%%%%%%%%%%%%
\label{average_estimate}
%%%%%%%%%%%%%%%%%%%%%%%%%%%%%%%%%%%%%%%%%%%%%%%%%%
\bar{V}_{\rm dir}/t \approx 9.37,
\quad
\bar{V}_{\rm um}/t \approx 8.93.
\end{eqnarray}
Note that, within our model, the two coupling constants are very close to
each other.

With such a replacement, the mean field interaction Hamiltonian can be
approximated as
\begin{equation}
%%%%%%%%%%%%%%%%%%%%%%%%%%%%%%
\label{MeanF_int2}
%%%%%%%%%%%%%%%%%%%%%%%%%%%%%%
\hat{H}_{\textrm{int}}^{\textrm{MF}}
=
-\sum_\mathbf{q}
	\textrm{Tr}\!\left(
		\hat{Q}^\dag\hat{\Theta}_\mathbf{q}
	       	+	
		\hat{\Theta}^\dag_\mathbf{q}\hat{Q}
	\right),
\end{equation}
where we introduced a matrix order parameter
$\hat{Q}$
\begin{equation}
%%%%%%%%%%%%%%%%%%%%%%%%%%%%%%
\label{OrderPar}
%%%%%%%%%%%%%%%%%%%%%%%%%%%%%%
\hat{Q}
=
\frac{1}{2N_c}\sum_\mathbf{p}
	\left(
		\bar{V}_{\rm dir} \langle\hat{\Theta}_\mathbf{p}\rangle
		+
		\bar{V}_{\rm um} \langle\hat{\Theta}^\dag_\mathbf{p}\rangle
	\right),
\end{equation}
which is the $\mathbf{q}$ independent in the range
$|\mathbf{q}|<q_0$,
and equal to zero otherwise. The definition
Eq.~\eqref{OrderPar}
can be inverted as
\begin{equation}
%%%%%%%%%%%%%%%%%%%%%%%%%%%%%%
\label{OrderPar_inv}
%%%%%%%%%%%%%%%%%%%%%%%%%%%%%%
\frac{1}{2N_c}\sum_\mathbf{p}\langle\hat{\Theta}_\mathbf{p}\rangle
=
\frac{1}{\bar{V}_{\rm dir}^2 - \bar{V}_{\rm um}^2}
\left(\bar{V}_{\rm dir}\hat{Q} - \bar{V}_{\rm um}\hat{Q}^\dag\right).
\end{equation}
In so doing, we have reduced formally the mean-field problem for AB-BLG to
the case considered in
Ref.~\onlinecite{aa2023su4_rozhkovetal}
for the AA-BLG. Thus, we can use the relation (60) from that paper
\begin{equation}
%%%%%%%%%%%%%%%%%%%%%%%%%%%%%%
\label{AA_59}
%%%%%%%%%%%%%%%%%%%%%%%%%%%%%%
\sum_{\mathbf{p}} \langle\hat{\Theta}_\mathbf{p}\rangle
=
\frac{1}{2}\sum_{\mathbf{q}}
	\hat{Q}
	\left(\varepsilon_0^2(\mathbf{q})+\hat{Q}^\dag\hat{Q}\right)^{-1/2}
\end{equation}
and derive a self-consistency equation in a convenient form
\begin{equation}
%%%%%%%%%%%%%%%%%%%%%%%%%%%%%%
\label{self-consist}
%%%%%%%%%%%%%%%%%%%%%%%%%%%%%%
\left(\bar{V}_{\rm dir}^2 -\bar{V}_{\rm um}^2 \right)
\hat{Q}h(\hat{Q}^\dag\hat{Q})
=
\bar{V}_{\rm dir}\hat{Q} - \bar{V}_{\rm um}\hat{Q}^\dag,
\end{equation}
where
\begin{equation}
%%%%%%%%%%%%%%%%%%%%%%%%%%%%%%
\label{hQQ}
%%%%%%%%%%%%%%%%%%%%%%%%%%%%%%
h(\hat{Q}^\dag\hat{Q})
=
\frac{1}{4N_c}\sum_{\mathbf{q}}
	\left(\varepsilon_0^2(\mathbf{q})+\hat{Q}^\dag\hat{Q}\right)^{-1/2}\!\!\!\!\!\!.
\end{equation}
%%%%%%%%%%%%%%%%%%%%%%%%%%%%%%%%%%%%%%%%%%%%%%%%%%%%%%%%%%%%%%%%%%%%%%%%%%%%%%%%%%%%%%%%%%%%%%%%%%%%%%%%%%%%%%%%%%%%%%%%%
The matrix
$\hat{Q}^\dag\hat{Q}$
is a non-negative Hermitian matrix. We use Hermitian-matrix representation
$\hat{Q}^\dag\hat{Q} = \hat{U} \hat{D}^2 \hat{U}^\dag$,
where
$\hat{U}$ is a unitary matrix and
\begin{eqnarray}
\hat{D} = {\rm diag}\, (d_1, \ldots, d_4),
\quad
d_i \geq 0,
\end{eqnarray}
is a diagonal matrix with non-negative eigenvalues. As a result, we can
write
\begin{eqnarray}
h(\hat{Q}^\dag\hat{Q})
=
\hat{U} h(\hat{D}^2) \hat{U}^\dag,
\end{eqnarray}
and
\begin{eqnarray}
h(\hat{D}^2)
=
\frac{\nu_0 }{4}
\int_0^{t_0} d \varepsilon \left(\varepsilon^2+\hat{D}^2\right)^{-1/2}
\approx
\frac{\nu_0 }{4}
\log \left( \frac{2t_0}{\hat{D}} \right),
\end{eqnarray}
where
$\nu_0 = t_0/(2\sqrt{3}\pi t^2)$
is the density of states (per valley per spin projection) in the undoped
AB-BLG at zero energy.

\section{Mean-field states}
%%%%%%%%%%%%%%%%%%%%%%%%%%%%%%%%%%%%%%%%%%%%%%%%%%
\label{sect::MF_states}
%%%%%%%%%%%%%%%%%%%%%%%%%%%%%%%%%%%%%%%%%%%%%%%%%%

The mean-field ordered phases of our SU(4)-symmetric Hamiltonian
are the solutions of the self-consistency
equation~(\ref{self-consist}).
It is quite remarkable that the same equation was derived in
Ref.~\onlinecite{aa2023su4_rozhkovetal}
for the different type of bilayer, AA-BLG [see Eq.~(61) in the latter
reference].

While the full set of solutions for
Eq.~(\ref{self-consist})
is unknown, we can identify several useful subsets. Let us assume that the
order parameter matrix
$\hat{Q}$
is normal (unitary-diagonalizable), that is, it can be represented as
\begin{eqnarray}
%%%%%%%%%%%%%%%%%%%%%%%%%%%%%%%%%%%%%%%%%%%%%%%%%%
\label{Q_general}
%%%%%%%%%%%%%%%%%%%%%%%%%%%%%%%%%%%%%%%%%%%%%%%%%%
\hat{Q} = \hat{U} {\rm diag}\, (q_1, q_2, q_3, q_4) \hat{U}^\dag,
\end{eqnarray}
where $\hat{U}$ is unitary, and
$q_j$
are complex
$q_j = |q_j|e^{i \phi_j} \in \mathbb{C}$.

Substituting such a matrix into the self-consistency equation, one can
establish that
$e^{i\phi_j}$
is either
$\pm 1$
or
$\pm i$.
This means that any
$q_j$
is either purely real, or purely imaginary. For
$|q_j|$
we derive
\begin{eqnarray}
\frac{\nu_0 }{4} {|q_j|}\log \left( \frac{2t_0}{|q_j|} \right)
=
\frac{|q_j|}{\bar{V}_{\rm dir} \pm \bar{V}_{\rm um}}\,,
\end{eqnarray}
such that one must choose plus sign (minus sign) for purely real (purely
imaginary) eigenvalue
$q_j$.

The latter equation is easy to solve and find that either
$q_j = 0$,
or
\begin{eqnarray}
%%%%%%%%%%%%%%%%%%%%%%%%%%%%%%%%%%%%%%%%%%%%%%%%%%
\label{q_solutions}
%%%%%%%%%%%%%%%%%%%%%%%%%%%%%%%%%%%%%%%%%%%%%%%%%%
q_j = \pm \Delta_+,
\quad
\text{or}
\
q_j = \pm i \Delta_-,
\end{eqnarray}
where
\begin{eqnarray}
%%%%%%%%%%%%%%%%%%%%%%%%%%%%%%%%%%%%%%%%%%%%%%%%%%
\label{Delta_pm_def}
%%%%%%%%%%%%%%%%%%%%%%%%%%%%%%%%%%%%%%%%%%%%%%%%%%
\Delta_\pm
=
2t_0 e^{-2/\lambda_\pm},
\end{eqnarray}
and we define two dimensionless coupling constants as
\begin{eqnarray}
%%%%%%%%%%%%%%%%%%%%%%%%%%%%%%%%%%%%%%%%%%%%%%%%%%
\label{lambda_pm_def}
%%%%%%%%%%%%%%%%%%%%%%%%%%%%%%%%%%%%%%%%%%%%%%%%%%
\lambda_\pm = \frac{\nu_0}{2}(\bar{V}_{\rm dir} \pm \bar{V}_{\rm um}).
\end{eqnarray}

One can prove that, if
$\hat{Q}$
is a solution of
Eq.~(\ref{self-consist}),
then
$\hat{V}\hat{Q}\hat{V}^\dag$ is also a solution
if $\hat{V} \in$\,SU(4).
In other words, all matrices that are unitary equivalent to a known
solution are solutions themselves. The inverse statement is not true: two
solutions to
Eq.~(\ref{self-consist})
are not necessarily unitary equivalent. For example, if
$\hat{Q}_1$
has one imaginary eigenvalue, while
$\hat{Q}_2$
has none, the two cannot be connected by a unitary transformation.
Consequently, the whole subset of mean field order parameters described by
Eq.~(\ref{Q_general})
can be split into several SU(4)-invariant multiplets.

Two unitary diagonalizable matrices are unitary equivalent only when their
eigenvalue sets are identical up to a permutation. This allows us to split
all
solutions~(\ref{Q_general})
into three sets according to how many real eigenvalues
$q_j$
a matrix has: (i)~all four
$q_j$
are real, (ii)~all
$q_j$
are imaginary. In set~(iii) some eigenvalues are real, some are purely
imaginary.

Within each of the three sets there are even finer divisions. To
illustrate this let us examine set~(i) with all real
$q_j$.
When all
$q_j$
are real, one can use
Eqs.~(\ref{Q_general})
and~(\ref{q_solutions})
and write
\begin{eqnarray}
%%%%%%%%%%%%%%%%%%%%%%%%%%%%%%%%%%%%%%%%%%%%%%%%%%
\label{Q_i}
%%%%%%%%%%%%%%%%%%%%%%%%%%%%%%%%%%%%%%%%%%%%%%%%%%
\hat{Q}
=
\Delta_+ \hat{U} {\rm diag}\, (\pm 1, \pm 1, \pm 1, \pm 1) \hat{U}^\dag.
\end{eqnarray}
In this expression all four signs are independent of each other and can be
chosen arbitrarily.

For matrices of this type, the invariant quantity is the number of positive
eigenvalues (up to overall sign change
$\hat{Q} \rightarrow -\hat{Q}$).
Thus, three classes of such matrices can be
introduced~\cite{Nandkishore2010b}
\begin{eqnarray}
%%%%%%%%%%%%%%%%%%%%%%%%%%%%%%%%%%%%%%%%%%%%%%%%%%
\label{eq::classI}
%%%%%%%%%%%%%%%%%%%%%%%%%%%%%%%%%%%%%%%%%%%%%%%%%%
\text{Class I:}&
\Delta_+ %\hat{U}
{\rm diag}\,(\kappa, \kappa, \kappa, \kappa),% \hat{U}^\dag,
&
\kappa = \pm 1,
\\
%%%%%%%%%%%%%%%%%%%%%%%%%%%%%%%%%%%%%%%%%%%%%%%%%%
\label{eq::classII}
%%%%%%%%%%%%%%%%%%%%%%%%%%%%%%%%%%%%%%%%%%%%%%%%%%
\text{Class II:}&
\Delta_+ \hat{U}
{\rm diag}\,(1,1,-1,-1) \hat{U}^\dag,
\\
%%%%%%%%%%%%%%%%%%%%%%%%%%%%%%%%%%%%%%%%%%%%%%%%%%
\label{eq::classIII}
%%%%%%%%%%%%%%%%%%%%%%%%%%%%%%%%%%%%%%%%%%%%%%%%%%
\text{Class III:}&
\Delta_+ \hat{U}
{\rm diag}\,(\kappa, \kappa, \kappa, -\kappa)\hat{U}^\dag,
&
\kappa = \pm 1.
\end{eqnarray}
Observe that, since class~I order parameter is proportional to unity
matrix, any unitary transformation leaves this order parameter unchanged.
Within this formalism, Kekul{\'{e}}-type order
parameters~\cite{chamon2000kekule_dist, cvetkovich2012bilayer_orders_theor}
(those that couple fermion states in non-identical valleys) are included
into these groups on equal footing with other order parameters types.

The order parameters from set~(i) represent various types of density waves
(charge-, spin-, \mbox{valley-,} spin-valley-density wave). To identify
these orders, we can use
Eq.~(\ref{d_q0LoEn_g})
to derive a relation
\begin{eqnarray}
\sum_{\bf q}
	\langle
		\gamma^\dag_{{\bf q} 3 \xi \sigma}
		\gamma^{\vphantom{\dag}}_{{\bf q} 2 \xi \sigma}
	\rangle
=
- \frac{\xi}{2}
\left(
	\chi_{ \xi \sigma} + i J_{ \xi\sigma}
\right),
\end{eqnarray}
where inter-layer partial polarizations
$\chi_m$
and dimensionless inter-layer ``currents''
$J_m$
are defined according to
\begin{eqnarray}
%%%%%%%%%%%%%%%%%%%%%%%%%%%%%%%%%%%%%%%%%%%%%%%%%%
\label{polarization_definition}
%%%%%%%%%%%%%%%%%%%%%%%%%%%%%%%%%%%%%%%%%%%%%%%%%%
\chi_{ \xi \sigma}
=
\sum_{\bf q}
	\langle
		d^\dag_{{\bf q} 1A \xi \sigma}
		d^{\vphantom{\dagger}}_{{\bf q} 1A \xi \sigma}
	\rangle
	-
	\langle
		d^\dag_{{\bf q} 2B \xi \sigma}
		d^{\vphantom{\dagger}}_{{\bf q} 2B \xi \sigma}
	\rangle,
\\
%%%%%%%%%%%%%%%%%%%%%%%%%%%%%%%%%%%%%%%%%%%%%%%%%%
\label{Jm_definition}
%%%%%%%%%%%%%%%%%%%%%%%%%%%%%%%%%%%%%%%%%%%%%%%%%%
J_{ \xi \sigma}
=
i \sum_{\bf q}
	e^{2 i \xi \phi_{\bf q} }
	\langle
		d^\dag_{{\bf q} 1A \xi \sigma }
		d^{\vphantom{\dagger}}_{{\bf q} 2B \xi \sigma }
	\rangle
	+ {\rm C.c.}
\end{eqnarray}
Both
$\chi_m$
and
$J_m$
are real.

On the other hand,
Eq.~(\ref{OrderPar_inv})
allows one to connect the sums
$\sum_{\bf q}
	\langle
		\gamma^\dag_{{\bf q} 3 m}
		\gamma^{\vphantom{\dag}}_{{\bf q} 2 m}
	\rangle$
to diagonal elements
${Q}_{m,m}$
of the order parameter matrix
$\hat{Q}$.
Thus, we derive
\begin{eqnarray}
%%%%%%%%%%%%%%%%%%%%%%%%%%%%%%%%%%%%%%%%%%%%%%%%%%
\label{ordered_polarizations}
%%%%%%%%%%%%%%%%%%%%%%%%%%%%%%%%%%%%%%%%%%%%%%%%%%
\chi_{\xi \sigma}
=
-
\frac{4 N_c \xi}{\bar{V}_{\rm dir} + \bar{V}_{\rm um}}
{\rm Re}\/ \left[ {Q}_{\xi \sigma, \xi \sigma} \right],
\\
%%%%%%%%%%%%%%%%%%%%%%%%%%%%%%%%%%%%%%%%%%%%%%%%%%
\label{ordered_currents}
%%%%%%%%%%%%%%%%%%%%%%%%%%%%%%%%%%%%%%%%%%%%%%%%%%
J_{\xi \sigma}
=
-
\frac{4 N_c \xi}{\bar{V}_{\rm dir} - \bar{V}_{\rm um}}
{\rm Im}\/ \left[ {Q}_{\xi \sigma, \xi \sigma} \right].
\end{eqnarray}
If matrix
$\hat Q$
is a diagonal representative of classes~I, II, or III, its diagonal matrix
elements are all real, as
Eqs.~(\ref{eq::classI}),
(\ref{eq::classII}),
and~(\ref{eq::classIII})
indicate. In this case the inter-layer ``currents'' identically vanish
$J_m \equiv 0$,
while all partial polarizations are finite
$|\chi_m| \!=\! 4 N_c \Delta_+/(\bar{V}_{\rm dir} + \bar{V}_{\rm um})$.
For diagonal order parameters in
Eqs.~(\ref{eq::classI})
(\ref{eq::classII})
and~(\ref{eq::classIII})
one can explicitly write
\begin{eqnarray}
%%%%%%%%%%%%%%%%%%%%%%%%%%%%%%%%%%%%%%%%%%%%%%%%%%
\label{eq::OP_diag}
%%%%%%%%%%%%%%%%%%%%%%%%%%%%%%%%%%%%%%%%%%%%%%%%%% 
Q_{\xi \sigma, \xi \sigma}
&=&
- \xi \frac{\bar{V}_{\rm dir} + \bar{V}_{\rm um}}{4 N_c }
\times
\\
\nonumber 
&&\sum_{\bf q}
	\langle
		d^\dag_{{\bf q} 1A \xi \sigma}
		d^{\vphantom{\dagger}}_{{\bf q} 1A \xi \sigma}
	\rangle
	-
	\langle
		d^\dag_{{\bf q} 2B \xi \sigma}
		d^{\vphantom{\dagger}}_{{\bf q} 2B \xi \sigma}
	\rangle,
\end{eqnarray} 
while all other matrix elements of
$\hat{Q}$
are zero.

For class~I state we obtain
$\chi_{\xi \sigma}
\!=\!
- 4 N_c \xi \kappa \Delta_+/(\bar{V}_{\rm dir} + \bar{V}_{\rm um})$.
This allows us to establish for such a state that the sample's inter-layer
electric polarization
$\chi_c$
\begin{eqnarray}
%%%%%%%%%%%%%%%%%%%%%%%%%%%%%%%%%%%%%%%%%%%%%%%%%%
\label{full_polarization}
%%%%%%%%%%%%%%%%%%%%%%%%%%%%%%%%%%%%%%%%%%%%%%%%%%
\chi_c = \sum_m \chi_m =
\langle \hat{\chi}_{\bf k} \rangle \Big|_{{\bf k} = 0}
\end{eqnarray}
vanishes. Spin and spin-valley inter-layer polarizations
\begin{eqnarray}
\chi_s = \sum_{\xi \sigma} \sigma \chi_{\xi \sigma},
\quad
\chi_{sv} = \sum_{\xi \sigma} \xi \sigma \chi_{\xi \sigma}
\end{eqnarray}
vanish as well. However, inter-layer valley polarization
\begin{eqnarray}
%%%%%%%%%%%%%%%%%%%%%%%%%%%%%%%%%%%%%%%%%%%%%%%%%%
\label{valley_polarization}
%%%%%%%%%%%%%%%%%%%%%%%%%%%%%%%%%%%%%%%%%%%%%%%%%%
\chi_v = \sum_{\xi \sigma} \xi \chi_{\xi \sigma}
%=
%-
%\frac{4 N_c}{\bar{V}_{\rm dir} + \bar{V}_{\rm um}}
%{\rm Re}\/ \left( {\rm Tr}\, \hat{Q} \right),
\end{eqnarray}
is finite
$\chi_v = - 4 \kappa \chi_0$,
where
\begin{eqnarray}
\chi_0
= 4 N_c \frac{\Delta_+}{\bar{V}_{\rm dir} + \bar{V}_{\rm um}}
\end{eqnarray}
is a ``polarization quantum". Note that in the
literature~\cite{zhand2011hall_state_classific}
this state is often referred to as quantum anomalous Hall (QAH) state.

Applying this reasoning to class~II states, one establishes that this class
hosts ferroelectric state
($Q_{\xi \sigma, \xi \sigma} \propto \xi$),
layer antiferromagnet state
($Q_{\xi \sigma, \xi \sigma} \propto \xi \sigma$),
and spin-valley-polarized state
($Q_{\xi \sigma, \xi \sigma} \propto \sigma$),
as well as various non-diagonal combinations of these orders. Ferroelectric
state is also
known~\cite{zhand2011hall_state_classific}
as quantum valley Hall (QVH), spin-valley-polarized state as quantum spin
Hall (QSH) state.

The ferroelectric state features non-zero inter-layer charge polarization
$|\chi_c| = 4 \chi_0$.
Three other polarizations are nullified.

In the layer antiferromagnet state (also known as spin-density wave) the
only finite polarization is
$|\chi_s| = 4\chi_0$.
Total spin polarization of the whole sample vanishes in this state, while
spin polarizations of individual layers are finite and opposite to each
other. Spin-valley-polarized state demonstrates similar characteristics
with respect to spin-valley polarization.

Class~III hosts eight diagonal
$\hat{Q}$,
all connected to each other by permutations of the eigenvalues in
definition~(\ref{eq::classIII}),
and by transformation
$\kappa \rightarrow -\kappa$.
It is easy to check that every diagonal order parameter represents a phase
with finite electric, valley, spin, and spin-valley inter-layer
polarizations that satisfy the following relation
\begin{eqnarray}
%%%%%%%%%%%%%%%%%%%%%%%%%%%%%%%%%%%%%%%%%%%%%%%%%%
\label{classIII_polarizations}
%%%%%%%%%%%%%%%%%%%%%%%%%%%%%%%%%%%%%%%%%%%%%%%%%%
|\chi_c| = |\chi_s| = |\chi_v| = |\chi_{sv}| = 2 \chi_0.
\end{eqnarray}
Unlike the absolute values, which are identical for every state with
diagonal
$\hat{Q}$,
the signs of the polarizations do depend on ordered phase. Direct
calculations of the polarizations demonstrate that for any
$\hat{Q}$-diagonal
phase the product of all polarizations is negative
\begin{eqnarray}
%%%%%%%%%%%%%%%%%%%%%%%%%%%%%%%%%%%%%%%%%%%%%%%%%%
\label{set_iii_coherence}
%%%%%%%%%%%%%%%%%%%%%%%%%%%%%%%%%%%%%%%%%%%%%%%%%%
\chi_c \chi_s \chi_v \chi_{sv} < 0.
\end{eqnarray}
In other words, either one polarization is negative, or three polarizations
are negative. This constraint allows for eight possible signs assignments,
each assignment represents a class-III phase with diagonal
$\hat{Q}$.

We can see that any class~III state combines within itself all
characteristics of the class~I and class~II states. Indeed, any class~III
state possesses finite charge, spin, valley, and spin-valley polarizations.
At the same time, in class~III the polarizations absolute value is
$2 \chi_0$
which is two times smaller than polarization values in the classes~I and~II.

In addition to polarizations, these three classes can be described in term
of quantized Hall resistance, see
Ref.~\onlinecite{Nandkishore2010b}.
According to the latter reference, both class~I and~III can be viewed as an
instance of quantum anomalous Hall state: they have finite (and quantized)
Hall conductance at zero magnetic field. Class~II phases have vanishing
electric Hall conductance, yet, they exhibit quantum flavor Hall effect.

Unlike set~(i), for set~(ii)
\begin{eqnarray}
%%%%%%%%%%%%%%%%%%%%%%%%%%%%%%%%%%%%%%%%%%%%%%%%%%
\label{Q_set_ii}
%%%%%%%%%%%%%%%%%%%%%%%%%%%%%%%%%%%%%%%%%%%%%%%%%%
\hat{Q}
=
\Delta_{-} \hat{U} {\rm diag}\, (\pm i, \pm i, \pm i, \pm i) \hat{U}^\dag,
\end{eqnarray}
all diagonal order parameters are characterized by vanishing polarizations
$\chi_m \equiv 0$.
At the same time, 
$J_m$
are finite, as
Eq.~(\ref{ordered_currents})
indicates. Depending on the assignment of signs in
Eq.~(\ref{Q_set_ii}),
corresponding inter-layer coherences are associated with certain quanta:
electric charge, spin, valley, spin-valley, or some combination thereof.

Alternatively, one can view such sates as a inter-layer-exciton-type order.
To explain this, consider
Eq.~(\ref{Jm_definition}).
We see that non-zero
$J_{\xi \sigma}$
implies that
$\langle
	d_{{\bf q} 2B \xi \sigma}^\dag
	d_{{\bf q} 1A \xi \sigma}^{\vphantom{\dag}}
\rangle$
and
$\langle
	d_{{\bf q} 1A \xi \sigma}^\dag
	d_{{\bf q} 2B \xi \sigma}^{\vphantom{\dag}}
\rangle$
are non-zero as well. The structure of these averages may be interpreted as
a type of symmetry-breaking inter-layer coupling between a particle in one
layer and a hole in the other layer, that is, an inter-layer exciton.
Curiously, while
$J_{\xi \sigma}$
becomes finite only when a symmetry is broken, the matrix elements
$\langle
	d_{{\bf q} 2B \xi \sigma}^\dag
	d_{{\bf q} 1A \xi \sigma}^{\vphantom{\dag}}
\rangle$
and
$\langle
	d_{{\bf q} 1A \xi \sigma}^\dag
	d_{{\bf q} 2B \xi \sigma}^{\vphantom{\dag}}
\rangle$
are finite even when all symmetries are preserved, as direct calculation
demonstrates. For a diagonal order parameter in set~(ii) its non-zero
elements are
\begin{eqnarray}
%%%%%%%%%%%%%%%%%%%%%%%%%%%%%%%%%%%%%%%%%%%%%%%%%%
\label{eq::OP_diag_Im}
%%%%%%%%%%%%%%%%%%%%%%%%%%%%%%%%%%%%%%%%%%%%%%%%%%
Q_{\xi \sigma, \xi \sigma} = 
\xi \frac{\bar{V}_{\rm dir} - \bar{V}_{\rm um}}{4 N_c }
 \sum_{\bf q}
	e^{2 i \xi \phi_{\bf q} }
	\langle
		d^\dag_{{\bf q} 1A \xi \sigma }
		d^{\vphantom{\dagger}}_{{\bf q} 2B \xi \sigma }
	\rangle
-
\\
\nonumber 
	e^{- 2 i \xi \phi_{\bf q} }
	\langle
		d^\dag_{{\bf q} 2B \xi \sigma }
		d^{\vphantom{\dag}}_{{\bf q} 1A \xi \sigma }
	\rangle.
\end{eqnarray} 
Clearly, all other elements are zero. This expression is analogous
to 
Eq.~(\ref{eq::OP_diag})
for set~(i).

Matrices defined by
Eq.~(\ref{Q_set_ii})
can be classified according to their signatures [how many pluses and
minuses are in
Eq.~(\ref{Q_set_ii})].
In complete analogy to the classification we have outlined above for
set~(i), three classes of matrices can be identified for set~(ii) as well.

Finally, there is the set~(iii) of the order parameters matrices. It can be
presented as
\begin{eqnarray}
%%%%%%%%%%%%%%%%%%%%%%%%%%%%%%%%%%%%%%%%%%%%%%%%%%
\label{set_iii}
%%%%%%%%%%%%%%%%%%%%%%%%%%%%%%%%%%%%%%%%%%%%%%%%%%
\hat{Q}
=
\hat{U} {\rm diag} (\pm i \Delta_-, q_2, q_3, \pm \Delta_+)\hat{U}^\dag,
\end{eqnarray}
where
$q_2$
can be either real
$q_2 = \pm \Delta_+$,
or imaginary
$q_2 = \pm i \Delta_-$.
The same is true for
$q_3$.
As we can see, among four eigenvalues of
$\hat{Q}$
there is at least one that is purely real, and at least one that is purely
imaginary. This condition guarantees that
Eq.~(\ref{set_iii})
defines a separate kind of order parameters, which cannot be reduced to two
previously defined sets. It is not difficult to generalize
Eqs.~(\ref{eq::OP_diag})
and~(\ref{eq::OP_diag_Im})
for set~(iii): if
$Q_{\xi \sigma, \xi \sigma}$
is purely real, 
Eq.~(\ref{eq::OP_diag})
should be used. Otherwise, 
Eq.~(\ref{eq::OP_diag_Im})
is appropriate.

\begin{table}[ht]
\centering
\begin{tabular}{||c|c|c||}
\hline
1 real eigenvalue &
2 real eigenvalues &
3 real eigenvalues \\
$(\pm 1, \pm i, \pm i, \pm i)$&
$(\pm 1, \pm 1, \pm i, \pm i)$ &
$(\pm 1, \pm 1, \pm 1, \pm i)$ \\
\hline
\hline
$(\kappa, i\zeta, i\zeta, i\zeta)$&
$(\kappa, \kappa, i\zeta, i\zeta)$&
$(\kappa, \kappa, \kappa, i\zeta)$\\
\hline
$(\kappa, i\zeta, i, -i)$&
$(\kappa, \kappa, i, -i)$&
$(\kappa, 1, -1, i\zeta)$\\
\hline
& $(1, -1, i\zeta, i\zeta)$& \\
\hline
& $(1, -1, i, -i)$& \\
\hline
\hline
\end{tabular}
\caption{Classification of set~(iii) order parameters. All matrices
$\hat{Q}$
described by
Eq.~(\ref{set_iii})
can be split into three types according to the number of real eigenvalues.
Each column of the table represents one of of these types. We use the
following notations:
$\pm 1$
stands for a real eigenvalue,
$\pm i$
stands for  an imaginary eigenvalue, and indices $\kappa$ and $\zeta$ are
$\kappa = \pm 1$
and
$\zeta = \pm 1$.
Within each type, additional sub-types can be defined, according to the
number of minus signs in front of real and imaginary eigenvalues (up to a
permutation of eigenvalues).
%%%%%%%%%%%%%%%%%%%%%%%%%%%%%%%%%%%%%%%%%%%%%%%%%%
\label{table::classification}
%%%%%%%%%%%%%%%%%%%%%%%%%%%%%%%%%%%%%%%%%%%%%%%%%%
}
\end{table}

The order parameters of set~(iii) can be split in three groups according to
the number of real eigenvalues: the first group represents
$\hat{Q}$'s
with one real eigenvalue, the second group is for
$\hat{Q}$'s
with two, and the third group is for order parameters with three real
eigenvalues. Within each group smaller subdivisions can be introduced, see
Table~\ref{table::classification}.
The order parameters in set~(iii) shares the features of both density wave
states and inter-layer-exciton states.

\section{Discussion}
%%%%%%%%%%%%%%%%%%%%%%%%%%%%%%%%%%%%%%%%%%%%%%%%%%
\label{sect::discuss}
%%%%%%%%%%%%%%%%%%%%%%%%%%%%%%%%%%%%%%%%%%%%%%%%%%

Here we investigated the SU(4)-symmetric model of AB-BLG. Quite remarkably,
the obtained self-consistency equation coincides formally with that for the
AA-BLG [see Eq.~(61) from
Ref.~\onlinecite{aa2023su4_rozhkovetal}].
This allows us to use the order parameter classification scheme from the
latter paper.

Despite mathematical similarities in the self-consistency equations, there
are obvious differences between the two cases. Probably the most obvious
one is the difference between electron spectra: AA-BLG demonstrates linear
dispersion with a well-defined Fermi surface, while the dispersion is
quadratic, with Fermi points, for undoped AB-BLG. The order parameter
interpretations are not identical either.

Also, to derive the SU(4)-symmetric form of the interaction Hamiltonian, we
assumed that the dipole-dipole interaction and backscattering can be
neglected. Unlike this, for AA-BLG, only the scattering channels with large
transferred momenta (backscattering) were discarded to guarantee the
desired symmetry. As a result, the coupling constants entering the
self-consistency equations have different physical meanings for two
different types of bilayer.

The AB-BLG model with the SU(4) symmetry group was formulated in
Ref.~\onlinecite{Nandkishore2010b},
whose authors limited the scope to the Hermitian
$\hat{Q}$
only. This means that they discussed only the set~(i) solutions, and
ignored two other possibilities, sets~(ii) and~(iii). It is quite easy to
argue that if
$\bar V_{\rm dir}$
and
$\bar V_{\rm um}$
are identical, only Hermitian
$\hat{Q}$
are allowed. Indeed, if
$\bar V_{\rm dir} = \bar V_{\rm um}$,
we obtain from
Eq.~(\ref{OrderPar})
that
$\hat{Q}=\hat{Q}^\dag$,
that is, the order parameter belongs to set~(i), which we discussed in
Sec.~\ref{sect::MF_states}
in great details.

While in our model
$\bar V_{\rm dir}$
and
$\bar V_{\rm um}$
are unequal, they are quite close to each other, as
Eq.~(\ref{average_estimate})
illustrates. Consequently, the characteristic energy scale
$\Delta_+$
for set~(i) order parameters is much larger than
$\Delta_-$
for set~(ii). Thus, within the SU(4)-symmetric model with Coulomb
interaction, the ground state most probably corresponds to a solution from
set~(i).

At the same time, when momentum dependence of
$V^+_{\mathbf{q}-\mathbf{p}}$
is ignored, 
definitions~(\ref{V_sin_cos})
and~(\ref{average_interaction})
clearly indicate that the direct and umklapp coupling constants become
identical,
$\bar V_{\rm dir} = \bar V_{\rm um}$,
leading to disappearance of order parameters with imaginary eigenvalues,
which we grouped into sets~(ii) and~(iii). Such a contraction of order
parameter list may happen if interaction is approximated by local terms
before direct and umklapp contributions become identified as separate
entities. This seems to be the reason why the ordered states with imaginary
eigenvalues are absent from rather extensive order parameter lists compiled
in
Refs.~\onlinecite{cvetkovich2012bilayer_orders_theor, lemonic_rg_nemat_long2012},
see Eq.~(2.10) in
Ref.~\onlinecite{lemonic_rg_nemat_long2012},
and Eq.~(12) in
Ref.~\onlinecite{cvetkovich2012bilayer_orders_theor}.

The SU(4)-symmetric model offers a convenient framework for classification
of the ordered states, as we pointed out in our previous
paper~\cite{aa2023su4_rozhkovetal}
on AA-BLG. It allows to group numerous non-superconducting order parameters
compatible with the AB-BLG electronic structure in three sets, which can be
further divided into smaller subsets.

A remarkable success of this procedure is the discovery and identification
of  set~(ii) and~(iii) phases, whose existence was established as a result
of a purely mathematical analysis of the self-consistency condition.
Ordered states of this kind do not fit into an ordinary density-wave
typification. These examples demonstrate the utility of the SU(4)-symmetric
model as it allows one to recognize those orders that go beyond one's most
immediate and intuitive generalizations.

At the same time, the SU(4)-symmetric model results must be treated with
caution when experiment is discussed. Indeed, for AB-BLG, such a model
ignores several non-symmetric contributions, for example, the
$V^{-}$~term
in
Eq.~(\ref{Int_low_E2}).
Figure~\ref{interaction}
clearly shows that, while the dipole-dipole coupling constant
$V^{-}_{\bf k}$
is smaller than
$V^{+}_{\bf k}$,
it is not negligible.

Adding non-SU(4)-symmetric contributions (e.g., Hubbard-like on-site
repulsion, dipole-dipole interaction in
$\hat{H}_{\rm int}$),
or external perturbations (e.g., magnetic field, normal electric field), or
phonons may shift the balance between various ordered phases. Specifically,
degeneracy between the states within a given set will be lifted, at least
partially.

To illustrate this, consider the following term in the Hamiltonian
\begin{eqnarray}
	\hat{H}_E = \frac{1}{2} eEd \hat{\chi}.
\end{eqnarray}
It accounts for the effects of the transverse electric field $E$ arising
either due to applied gate voltage of spontaneous polarization. (Here
$d \sim 3$\,\AA\,
is the inter-layer distance.) The lowest-order correction due to this
term is
$\langle \hat{H}_E \rangle = eEd \chi_c/2$.
Since several ordered states in sets~(i) and~(iii) have finite
$\chi_c$,
we conclude that the external electric field partially destroys the
degeneracy between the solutions described in the previous section. Namely,
for set~(i), the energy of the ferroelectric states, depending on the
direction of ferroelectric polarization
$\chi_c$,
increases by
$2 e|E|d \chi_0$,
or decreases by the same amount. Other class~II and class~I states are not
affected by the field in this order of perturbation theory in powers of $E$.

Curiously,
$\hat{H}_E$
splits class~III into two subclasses, each with opposite sign of
$\chi_c = \pm 2 \chi_0$.
The energy of one subclass grows by
$e |E| d \chi_0$.
The energy of the other subclass decreases by the same amount.

Another interaction that destroys the SU(4) symmetry is an intrinsic
spin-orbit coupling (SOC), which is frequently discussed in literature.
However, a characteristic scale of the SOC energy in the AB-BLG is about
few meV or even less and corresponding energy splitting is estimated as
several tens of $\mu$eV (see, e.g.,
Refs.~\onlinecite{konschuh2012theory, guinea2010spin}). 
For comparison, a characteristic Coulomb interaction
energies scales are several eV or even larger, as revealed by
Fig.~\ref{interaction},
giving
rise~\cite{sboychakov2023triplet}
to order parameters values in the range of several meV.

Beside $E$ and SOC, yet another commonly studied type of non-symmetric
perturbations is the trigonal warping. However, corresponding first-order
corrections are zero. More complicated calculations, going beyond this
simple discussion, are required in this case.

A different, fluctuations-based, mechanism of the degeneracy
lifting was investigated in
Ref.~\onlinecite{Nandkishore2010b}
for set~(i) only. In a real material, several of these mechanisms function
simultaneously.

Inclusion of the new terms in the model Hamiltonian may affect the values
of the coupling constants. As we discuss in
Appendix~\ref{coupling_constants},
this could have very important consequences for theoretical description of
graphene bilayers.

Note also that we cannot include recently observed superconducting states
in our classification scheme. Applied gate voltage breaks the SU(4)
symmetry of the model. The superconducting states are observed in the
AB-BLG only when the gate voltage is
non-zero~\cite{zhou2022isospin}.
The same follows from the mean-field
considerations~\cite{sboychakov2023triplet}.
Thus, significant modifications must be introduced to our framework to
include superconducting phases.

To conclude, we investigated ordered phases of a SU(4)-invariant model for
the AB-BLG. Within the framework of the mean field approximation matrix
self-consistency equation was derived. It coincides with the
self-consistency equation obtained previously for AA-BLG, and demonstrates
rich diversity of solutions. Every solution represents a stationary
non-superconducting ordered phase of the model. Symmetry-based
classification of these orders is discussed.

\appendix

\section{Coupling constants}
%%%%%%%%%%%%%%%%%%%%%%%%%%%%%%%%%%%%%%%%%%%%%%%%%%
\label{coupling_constants}
%%%%%%%%%%%%%%%%%%%%%%%%%%%%%%%%%%%%%%%%%%%%%%%%%%

Here we discuss the values of the order parameters in the
SU(4)-symmetric model of the AB-BLG. Of course, this issue has been
addressed in literature. For example, the value of the order parameter was
estimated in
Ref.~\onlinecite{Nandkishore2010b}
as
\begin{eqnarray}
%%%%%%%%%%%%%%%%%%%%%%%%%%%%%%%%%%%%%%%%%%%%%%%%%%
\label{Delta_NL}
%%%%%%%%%%%%%%%%%%%%%%%%%%%%%%%%%%%%%%%%%%%%%%%%%%
\Delta = \Lambda  e^{-2/ \lambda \nu_0 },
\end{eqnarray}
see the formula beneath Eq.~(12) of the latter paper. In this expression
$\Lambda$ is the characteristic band energy
$\Lambda \sim t_0$,
and the dimensionless product
$\lambda \nu_0$ was
evaluated as
\begin{eqnarray}
%%%%%%%%%%%%%%%%%%%%%%%%%%%%%%%%%%%%%%%%%%%%%%%%%%
\label{lambda_NL}
%%%%%%%%%%%%%%%%%%%%%%%%%%%%%%%%%%%%%%%%%%%%%%%%%%
\lambda \nu_0 = \frac{1}{4 \ln 4} \approx 0.18.
\end{eqnarray}
This parameter is an analogue of our coupling constant
$\lambda_+$.
Equation~(\ref{lambda_NL})
is based on the RPA calculations from
Ref.~\onlinecite{Hwang2008}
with the polarization operator evaluated at zero momentum
${\bf q} = 0$
[see Eq.~(5) there]:
\begin{eqnarray}
\Pi^0 = \frac{2m}{\pi} \ln 4 = \frac{4t_0}{9 \pi t^2 a_0^2} \ln 4.
\end{eqnarray}
Here
$m=q_0^2/2t_0$
denotes an effective electron mass in AB-BLG. Substituting
\begin{eqnarray}
\Lambda = 0.35\,{\rm eV},
\quad
\frac{\lambda \nu_0}{2} = 0.09,
\end{eqnarray}
into~(\ref{Delta_NL}),
we obtain
$\Delta \sim 0.055$\,K.

We can recover this estimate if we make certain simplifications to our
framework. Following
Ref.~\onlinecite{Nandkishore2010b},
we ignore the momentum-dependence of the interaction
\begin{eqnarray}
V^+_{\bf q} \approx V^+_{\bf q} \Big|_{{\bf q} = 0}
\approx 26.3 t,
\end{eqnarray}
see
Fig.~\ref{interaction}.
Under this assumption
\begin{eqnarray}
%%%%%%%%%%%%%%%%%%%%%%%%%%%%%%%%%%%%%%%%%%%%%%%%%%
\label{coupling_estimate_approx}
%%%%%%%%%%%%%%%%%%%%%%%%%%%%%%%%%%%%%%%%%%%%%%%%%%
\bar V_{\rm dir} = \bar V_{\rm um}
=
\frac{1}{2} V^+_{\bf q} \Big|_{{\bf q} = 0} \approx 13.2t,
\quad
\Rightarrow
\quad
\lambda_+ \approx 0.16.
\quad
\end{eqnarray}
The consistency between the values of
$\lambda_+$
and
$\nu_0 \lambda$
from
Eq.~(\ref{lambda_NL})
is obvious. Using this estimate for
$\lambda_+$
and
Eq.~(\ref{Delta_pm_def}),
we obtain
$\Delta_+ \approx 0.026$\,K.

The calculated energy scales are very small and inconsistent with typical
transition temperatures between different states observed in the AB-BLG
samples~\cite{Velasco2012, veligura2012, Mayorov2011, Freitag20122053},
the latter is tens of Kelvin. The situation becomes more problematic if we
realize that the interaction is a decreasing function of the transferred
momentum, see
Fig.~\ref{interaction}.
Thus, the estimates in
Eq.~(\ref{coupling_estimate_approx})
are, in fact, upper bounds. If the momentum dependence of the interaction
is accounted for,
values~(\ref{average_estimate})
are derived. As expected, they are smaller than the values in
Eq.~(\ref{coupling_estimate_approx}),
which implies even smaller order parameter scale.

All this indicates that the SU(4)-symmetric Hamiltonian alone is not sufficient
for evaluation of the order parameters. For example, our previous RPA
study~\cite{sboychakov2023triplet}
of the non-SU(4) model for AB-BLG estimates the SDW order parameter as
$\Delta^{\rm SDW} = 4.9$\,meV,
in general consistency with the experimental data. Factors that increase the
order parameter in the non-SU(4)-symmetric model are the following. First,
in Ref.~\onlinecite{sboychakov2023triplet} we took into account the
$V^{-}_{\mathbf{q}}$ interaction term which increases coupling constant
$\lambda$. Since
$\lambda^{-1}$
is placed in the exponential function, even moderate growth of $\lambda$
can substantially enhance the transition temperature. Further, in
Ref.~\onlinecite{sboychakov2023triplet}
we used $4$-band model of the AB-BLG, and extend the momentum integration
in the self-consistency equation to a much larger area in reciprocal space.
Effectively this leads to higher pre-exponential factor in the formula for
the transition temperature.

Thus, while the SU(4) symmetry of a bilayer graphene model is a useful
classification tool, it underestimates order parameter strength.

\end{document}